\documentclass{amsart}
\usepackage{latexsym}
\usepackage{amsfonts}
\usepackage{amssymb}
\usepackage{graphicx}

\newtheorem{theorem}{Theorem}[section]

\theoremstyle{definition}

\newtheorem{fact}{Fact}[section]
\theoremstyle{remark}

\numberwithin{equation}{section}

\newcommand{\Lg}{\mathfrak{g}}
\newcommand{\Lh}{\mathfrak{h}}

\newcommand{\Lp}{\mathfrak{p}}
\newcommand{\La}{\mathfrak{a}}

\newcommand{\Lt}{\mathfrak{t}}

\newcommand{\Ad}{\mathrm{Ad}}

\newcommand{\z}{\mathrm{z}}

\begin{document}

\title[] {A Criterion for the Covering Condition of Generalized
Random Matrix Ensembles}

\author[]{Jinpeng An}
\address{School of mathematical science, Peking University,
 Beijing, 100871, P. R. China }
\email{anjinpeng@pku.edu.cn}

\author[]{Zhengdong Wang}
\address{School of mathematical science, Peking University,
 Beijing, 100871, P. R. China}
\email{zdwang@pku.edu.cn}

\thanks{This work is supported by the 973 Project
Foundation of China (\#TG1999075102).}



\begin{abstract}
In this paper we present a criterion for the covering condition of
the generalized random matrix ensemble, which enable us to verify
the covering condition for the seven classes of generalized random
matrix ensemble in an unified and simpler way.
\end{abstract}

\maketitle


\section{Introduction}

In \cite{AWY1,AWY2} the authors introduced the concept of
generalized random matrix ensemble, and presented a classification
scheme of it by means of Lie theory, namely linear ensemble,
nonlinear noncompact ensemble, compact ensemble, group ensemble,
algebra ensembles, pseudo-group ensemble, and pseudo-algebra
ensemble. The seven classes of generalized ensemble include all
classical random matrix ensembles(see \cite{CM,Du,Me}) and some
new ensembles, the joint density functions of which were derived
in an unified way. Various kinds of classical integration formulae
(Weyl integration formula for compact Lie groups, Harish-Chandra's
integration formulae for complex semisimiple Lie groups and real
reductive groups, integration formulae associated with symmetric
spaces of noncompact and compact types, and their linear versions)
were also deduced as corollaries of an integration formula
associated with the generalized random matrix ensemble. The
integration formula associated with generalized ensemble relies on
a covering condition for the generalized ensemble, which asserts
that a natural map associated with the generalized ensemble is a
finite sheeted covering map. Two criterions of the covering
condition were proved in \cite{AWY1}, and were all used in a
confusing way when the authors verified the covering condition for
various kinds of generalized ensembles. This drawback somewhat
conflicts to the goal of the authors in \cite{AWY1,AWY2} which was
to present an unified theory of random matrix ensembles. The goal
of this paper is to fix this drawback. We present a criterion for
the covering condition of the generalized random matrix ensemble,
which enable us to verify the covering condition for the seven
classes of generalized ensemble in an unified way, and the
verifications for all of the seven cases are simpler than that of
in \cite{AWY2}.

Suppose a Lie group $G$ acts on a Riemannian manifold $X$ by
$\sigma: G\times X\rightarrow X$, and suppose the induced
Riemannian measure $dx$ is $G$-invariant. Let $Y$ be a closed
submanifold of $X$ with the induced Riemannian measure $dy$, and
let $K=\{g\in G:\sigma_g(y)=y, \forall y\in Y\}$. Define the map
$\varphi:G/K\times Y\rightarrow X$ by
$\varphi([g],y)=\sigma_g(y)$. Let $X_\z\subset X$, $Y_\z\subset Y$
be closed subsets of measure zero in $X$ and $Y$, respectively.
Denote $X'=X\setminus X_\z$, $Y'=Y\setminus Y_\z$. Suppose
the following conditions hold.\\
(\emph{a}) \quad (\emph{Invariance condition}) \quad $X'=
\bigcup_{y\in Y'}O_y$.\\
(\emph{b}) \quad (\emph{Transversality condition}) \quad $T_yX=T_y
O_y\oplus T_yY$, \quad $\forall y\in Y'$.\\
(\emph{c}) \quad (\emph{Dimension condition}) \quad
$\dim G_y=\dim K, \quad \forall y\in Y'$.\\
(\emph{d}) \quad (\emph{Orthogonality condition}) \quad $T_yY\perp
T_yO_y$, \quad $\forall y\in Y'$.\\
(\emph{e}) \quad (\emph{Covering condition}) \quad \emph{There is
a positive integer $d$ such that the map $\varphi: G/K\times
Y'\rightarrow X'$ is a $d$-sheeted covering map.}\\
Suppose $d\mu$ is a $G$-invariant smooth measure on $G/K$, and
suppose $p(x)$ is a $G$-invariant smooth function on $X$. Then the
system $(G,\sigma,X,p(x)dx,Y,dy)$ is a generalized random matrix
ensemble. It is proved in \cite{AWY1} that there is a generalized
eigenvalue distribution $d\nu$ on $Y$ such that
$\varphi^*(p(x)dx)=d\mu d\nu$, and we have the integration formula
\begin{equation}\label{E:integration}
\int_X f(x)p(x)
dx=\frac{1}{d}\int_Y\left(\int_{G/K}f(\sigma_g(y))d\mu([g])\right)
d\nu(y)
\end{equation}
for all $f\in C^\infty(X)$ with $f\geq0$ or with $f\in
L^1(X,p(x)dx)$.

Let $N=\{g\in G:\sigma_g(Y)=Y\}$. Then $N$ is a closed subgroup of
$G$, $K$ is a normal subgroup of $N$. For $y\in Y$, denote
$N_y=\{g\in G:\sigma_g(y)\in Y\}$. The main result in this paper
is the following conclusion.

\begin{theorem}\label{T:main}
Suppose conditions (a), (b) and (c) hold. If $N/K$ is a finite
group of order $d$, and $N_y=N$ for every $y\in Y'$, then the
covering condition (e) holds, that is, $\varphi: G/K\times
Y'\rightarrow X'$ is a $d$-sheeted covering map.
\end{theorem}

Theorem \ref{T:main} will be proved in Section 2. Using Theorem
\ref{T:main}, the covering condition for the seven classes of
generalized ensemble will be re-verified in Section 3.


\section{Proof of Theorem \ref{T:main}}

We first recall some facts which were proved in \cite{AWY1}.

\begin{fact}\label{F:1}
(\cite{AWY1}, Proposition 3.2.) \emph{Suppose conditions (a), (b)
and (c) hold. Then $\varphi: G/K\times Y'\rightarrow X'$ is
everywhere regular.}
\end{fact}

\begin{fact}\label{F:2}
(\cite{AWY1}, Proposition 3.5.) \emph{Let $M, N$ be smooth
manifolds of the same dimension, $d$ a positive integer. Then an
everywhere regular smooth map $f:M\rightarrow N$ is a $d$-sheeted
covering map if and only if for every $p\in N$, $\varphi^{-1}(p)$
has $d$ points.}
\end{fact}

\begin{proof}[Proof of Theorem \ref{T:main}]
It can be easily verified from the dimension condition (c) that
$\dim(G/K\times Y')=\dim X'$ (see also \cite{AWY1}). By Facts
\ref{F:1} and \ref{F:2}, it is sufficient to prove that for every
$x\in X'$, $\varphi^{-1}(x)$ has $d$ points.

Let $x\in X'$. By the invariance condition (a), there is some
$g\in G$ such that $\sigma_g(x)\in Y'$. Denote $y=\sigma_g(x)$.
Since $\varphi^{-1}(y)=(l_g\times id)(\varphi^{-1}(x))$, where
$l_g$ is the left multiplication of $g$ on $G/K$, it is sufficient
to show $\varphi^{-1}(y)$ has $d$ points.

Now choose $g_1, \cdots, g_d\in N$, one in each coset space of $K$
in $N$. Then we have $\{([g_i], \sigma_{g_i^{-1}}(y)): i=1,
\cdots, d\}\subset\varphi^{-1}(y)$. On the other hand, if
$([g'],y')\in\varphi^{-1}(y)$, that is, $\sigma_{g'}(y')=y$, then
$g'\in N_{y'}=N$. So there is some $i_0\in\{1, \cdots, d\}$ such
that $[g']=[g_{i_0}]$, and then $([g'],y')=([g_{i_0}],
\sigma_{g_{i_0}^{-1}}(y))$. This shows that
$\varphi^{-1}(y)\subset\{([g_i], \sigma_{g_i^{-1}}(y)): i=1,
\cdots, d\}$. Hence $\varphi^{-1}(y)=\{([g_i],
\sigma_{g_i^{-1}}(y)): i=1, \cdots, d\}$, which has $d$ points.
\end{proof}


\section{Verification of the Covering Condition}

The covering condition was verified in \cite{AWY2} for the seven
classes of general ensembles respectively. One ingredient was
Corollary 3.6 in \cite{AWY1}, which said that if $|O_y\cap Y'|=d$
and the isotropic $G_y=K$ for every $y\in Y'$, then the covering
condition holds. But this criterion was not always available. For
example, when verifying the covering condition for the compact
ensemble and the group ensemble associated with complex semisimple
Lie groups, the authors had to appeal to a Proposition 3.5 in
\cite{AWY1}, that is Fact \ref{F:2} in this paper. In this section
we reverify the covering conditions for the seven classes of
generalized ensembles using Theorem \ref{T:main}. We will see that
the process of verification here is simpler than that of in
\cite{AWY2} for each of the seven cases.

\begin{theorem}\label{T:verify}
The linear ensemble, the nonlinear noncompact ensemble, the
compact ensemble, group ensembles associated with compact Lie
groups and complex semisimple Lie groups, algebra ensembles
associated with compact Lie groups and complex semisimple Lie
groups, the pseudo-group ensemble, and the pseudo-algebra ensemble
satisfy the covering condition.
\end{theorem}

\begin{proof}
We adopt the notations in corresponding sections of \cite{AWY2}.

\emph{(i) Linear ensemble.} Since by Proposition 7.32 in \cite{Kn}
the group $W=N_K(\La)/Z_K(\La)$ is finite, by Theorem
\ref{T:main}, it is sufficient to prove that for $\eta\in\La'$, if
$k\in K$ such that $\eta'=\Ad(k)(\eta)\in\La'$, then $k\in
N_K(\La)$. Since $\Ad(k)$ is an automorphism of $\Lg$,
$\Ad(k)(Z_\Lg(\eta))=Z_\Lg(\eta')$. But
$\eta,\eta'\in\La'=\La\setminus(\bigcup_{\lambda\in\Sigma^+}\ker\lambda)$
implies $Z_\Lg(\eta)=Z_\Lg(\eta')=\Lg_0$. So $\Ad(k)$ fixes
$\Lg_0$. But $\Lp$ is also fixed by $\Ad(k)$. This implies
$\Ad(k)$ fixes $\Lg_0\cap\Lp=\La$, that is, $k\in N_K(\La)$,
finishing the proof of this case.

\emph{(ii) Nonlinear noncompact ensemble.} Since
$\exp|_\Lp:\Lp\rightarrow P$ is a diffeomorphism and
$\exp(\La')=A'$ by definition, this case is equivalent to the case
of linear ensemble.

\emph{(iii) Compact ensemble.} Since
$A'=A\setminus(\bigcup_{\lambda\in\Sigma^+}\ker\vartheta_\lambda)$,
for $a\in A'$, we have $Z_\Lg(a)=\Lg_0$. Similar to she proof for
linear ensemble, if $\sigma_k(a)\in A'$ for some $k\in K, a\in
A'$, then $\Ad(k)$ fixes $\La$, that is, $k\in N_K(\La)=N_K(A)$.

\emph{(iv) Group ensemble and algebra ensemble associated with
compact Lie groups.} Let $G$ be a compact Lie group with Lie
algebra $\Lg$, $T$ a maximal torus with Lie algebra $\Lt$. As in
\cite{AWY2}, let
$T'=T\setminus(\bigcup_{\alpha\in\Delta}\ker\vartheta_\alpha)$,
$\Lt'=\Lt\setminus(\bigcup_{\alpha\in\Delta}\ker\alpha)$. Then for
$t\in T'$ and $\eta\in\Lt'$, $Z_\Lg(t)=Z_\Lg(\eta)=\Lt$. So if
$t\in T'$ and $g\in G$ such that $gtg^{-1}\in T'$, then
$\Ad(g)(\Lt)=\Lt$, that is, $g\in N_G(\Lt)$. Similarly, if
$\eta\in\Lt'$ and $g\in G$ such that $\Ad(g)(\eta)\in\Lt'$, then
$g\in N_G(\Lt)$. By Theorem \ref{T:main}, the associated ensembles
satisfy the covering condition with covering sheet $|W|$.

\emph{(v) Group ensemble and algebra ensemble associated with
complex semisimple Lie groups.} It is almost same as the proof of
case (iv).

\emph{(vi) Pseudo-group ensemble and pseudo-algebra ensemble.} Let
$G$ be a real reductive group with Lie algebra $\Lg$, $\Lh_j$ a
Cartan subalgebra of $\Lg$, $H_j=Z_G(\Lh_j)$ the associated Cartan
subgroup of $G$. As in \cite{AWY2}, let $\Lh_j'=\Lh_j\cap\Lg_r$,
$H_j'=H_j\cap G_r$. For the associated pseudo-algebra ensemble, if
$\eta\in\Lh_j'$ and $g\in G$ such that $\Ad(g)(\eta)\in\Lh_j'$,
then $\Ad(g)(\Lh_j)=\Lh_j$, due to the fact that
$Z_\Lg(\eta)=\Lh_j$. By Theorem \ref{T:main}, the covering
condition is satisfied with covering sheet
$|W_j|=|N_G(\Lh_j)/H_j|$. For the associated pseudo-algebra
ensemble, if $h\in H'$ and $g\in G$ such that $ghg^{-1}\in H'$,
then $\Ad(g)(\Lh_j)=\Lh_j$, due to the fact that $Z_\Lg(h)=\Lh_j$.
Hence $g\in N_G(H_j)$, and the covering condition is satisfied
with covering sheet $|N_G(H_j)/Z_G(H_j)|=|W_j|\cdot|H_j/Z(H_j)|$,
which is finite by Proposition 7.25 in \cite{Kn}.
\end{proof}


\begin{thebibliography}{9}

\bibitem{AWY1} An, J., Wang, Z., Yan, K., \textit{A Generalization
of random matrix ensemble I: general theory}, Pacific J. Math., to
appear, math-ph/0502020.

\bibitem{AWY2} An, J., Wang, Z., Yan, K., \textit{A Generalization
of random matrix ensemble II: concrete examples and integration
formulae}, preprint, math-ph/0502021.

\bibitem{CM} Caselle, M., Magnea, U.,
\textit{Random matrix theory and symmetric spaces}, Phys. Rep.
394, 41-156, 2004.

\bibitem{Du} Due\~{n}ez, E., \textit{Random matrix
ensembles associated to compact symmetric spaces}, Comm. Math.
Phys. 244, 29--61, 2004.

\bibitem{Kn} Knapp, A. W., \textit{Lie groups beyond an introduction},
2nd edition, Birkh\"auser, Boston, 2002.

\bibitem{Me} Mehta, M. L., \textit{Random matrices}, Academic
Press, San Diego, 1991.
\end{thebibliography}
\end{document}